\documentclass[12pt]{article}

\usepackage{amssymb,amsmath}

\usepackage{graphicx}

\newcommand{\be}{\begin{equation}}
\newcommand{\ee}{\end{equation}}

\newcommand{\dlt}{\delta}
\newcommand{\prt}{\partial}

\newcommand{\ba}{{\bf a}}

\newcommand{\bt}{\beta}
\newcommand{\vp}{\varphi}

\newcommand{\ra}{\rightarrow}

\newcommand{\om}{\omega}
\newcommand{\Om}{\Omega}

\newcommand{\dgr}{\dagger}

\newcommand{\Lbd}{\Lambda}

\newcommand{\cH}{{\cal H}}
\newcommand{\cM}{{\cal M}}
\newcommand{\cD}{{\cal D}}

\newcommand{\lgl}{\langle}
\newcommand{\rgl}{\rangle}

\begin{document}

\begin{center}

{\Large{\bf
Mesoscopic disorder in double-well optical lattices} \\ [5mm]
V.I. Yukalov$^1$ and E.P. Yukalova$^2$} \\ [3mm]

{\it $^1$ Bogolubov Laboratory of Theoretical Physics, \\
Joint Institute for Nuclear Research, Dubna 141980, Russia \\ [3mm]

$^2$ Laboratory of Information Technologies, \\
Joint Institute for Nuclear Research, Dubna 141980, Russia}
\end{center}

\vskip 3cm

\begin{abstract}

Double-well optical lattices are considered, each cite of which is
formed by a double-well potential. The lattice is assumed to be in
an insulating state and order and disorder are defined with respect
to the displacement of atoms inside the double-well potential. It
is shown that in such lattices, in addition to purely ordered and
disordered states, there can exist an intermediate mixed state,
where, inside a generally ordered lattice, there appear disordered
regions of mesoscopic size.
\end{abstract}

\vskip 5cm
{\bf E-mail}: yukalov@theor.jinr.ru

\newpage

\section {Introduction}

Cold atoms, loaded into optical lattices, form the systems that
are unique with regard to the controllability of their properties
(see the review articles [1-5] and references therein). One can
vary the sorts of atoms as well as the lattice types and lattice
spacing. For the given kind of atoms, one can vary their interactions
by means of the Feshbach resonance techniques [5-7]. It is also
possible to change the lattice properties by imposing external fields
or shaking the lattice. This is why, one is able to create a variety
of different system states in the lattice [1-5].

A special type of lattices are the double-well lattices, whose
experimental realization has become possible in recent years [8-12].
In such lattices, each lattice site is formed by a double-well
potential, which makes it feasible to create novel states that would
be impossible in the standard single-well lattices. The phase diagram
of bosons in double-well lattices has been studied [13,14], with the
emphasis on the boundary of the insulator-superfluid transition.
Double-well lattices in an insulator state have been analyzed [15-17],
with the emphasis on collective excitations in these lattices [15],
the order-disorder transition [15], dynamics of atomic loading in the
lattice [15,16], and the feasibility of regulating atomic imbalance by
varying the double-well potential shape [17]. The possibility of
transferring atoms between the wells of a double-well potential by
means of the stimulated Raman adiabatic passage scheme has been
suggested [18].

In the present paper, the possibility of creating another state in
a double-well optical lattice is advanced. We consider an insulating
double-well lattice, with a single atom per cite, which is a setup
that would be typical for the Mott insulator in a single-well lattice.
Because of the insulating state and the unity filling factor, the atom
statistics are not crucial; the atoms can be bosons as well as fermions.
In such a system, there can exist an {\it ordered} state, when all
atoms are unidirectionally shifted into one of the wells of each of
the double-well potentials [15]. Such a state is schematically shown
in Fig. 1. It may also happen a {\it disordered} state, when atoms are
randomly located in one of the wells of the double-well potential [15].
We analyze under what conditions there could arise an intermediate
mixed state, when inside a generally ordered lattice there could appear
mesoscopic regions of disordered states, as is illustrated in Fig. 2.
The term {\it mesoscopic} refers to the characteristic size $l_{mes}$
of the disordered regions, as compared to the lattice spacing $a$ and
to the total system size $L$. Of course, the disordered regions can
be of different sizes and shapes, but it is always possible to define
their characteristic, or average, size $l_{mes}$. This size is
mesoscopic in the sense of being intermediate between the lattice
spacing and system sizes:
$$
a \ll l_{mes} \ll L \;  .
$$

We show that there exist such system parameters, when the intermediate
mixed state of a lattice with mesoscopic disorder can arise. Since
optical lattices allow for extremely wide possibilities in varying
the system parameters, the suggested mixed state can be realized
experimentally. And, as far as the properties of the mixed state can
be regulated, it can be used for a variety of applications.

\section {Random Region Configurations}

In order to understand whether and when the mixed state could appear,
it is necessary to take into account its possible existence, so that
the probability of this state could be defined self-consistently. A
system, formed by subregions corresponding to different thermodynamic
states, having different orders, can be described by using the
local-equilibrium Gibbs ensembles [19,20]. The regions of competing
thermodynamics phases can arise in a self-organized way in arbitrary
spatial locations, that is, their locations are randomly distributed
in space. Therefore, it is necessary to average over these random
configurations. Such a procedure has been suggested in Refs. [21-27]
and reviewed in Ref. [28]. All mathematical details of this approach
has been thoroughly described in these references. To better understand
the following, in this section, we briefly delineate the main points
of the approach as applied to a lattice system.

Let us consider the lattice
\be
\label{1}
\mathbb{Z} = \left \{ \ba_i : \; i=1,2,\ldots, N_L \right \}
\ee
composed of $N_L$ lattice cites characterized by the vectors ${\bf a}_i$.
Suppose that there can exist two different thermodynamic states
corresponding to an ordered and disordered phases. Meanwhile, it is not
essential what kind of order is assumed. But for what follows, we keep
in mind that each lattice cite is formed by a double-well potential and
that the order is understood in the sense of the existence of an average
atomic imbalance, as is explained in the Introduction.

When a part of the lattice, say $\mathbb{Z}_1$, is filled by an ordered
state and a part $\mathbb{Z}_2$, by a disordered one, the whole lattice
is represented as the union
\be
\label{2}
\mathbb{Z} = \mathbb{Z}_1 \bigcup \mathbb{Z}_2 \qquad
\left (\mathbb{Z}_1 \bigcap \mathbb{Z}_2 = 0 \right ) \; ,
\ee
in which
$$
N_L = N_1 + N_2 \qquad ( N_\nu \equiv {\rm mes} \;
\mathbb{Z}_\nu ) \;  .
$$
This is what is called the {\it orthogonal covering}. Each manifold
covering can be characterized [29] by the manifold characteristic
functions, or {\it manifold indicators}
\begin{eqnarray}
\xi_{i\nu} = \left \{ \begin{array}{ll}
1 \; , & \ba_i \in \mathbb{Z}_\nu  \; ,\\
0 \; , & \ba_i \not\in \mathbb{Z}_\nu \; ,
\end{array} \right.
\end{eqnarray}
describing the lattice separation into sublattices. The collection of
the given manifold indicators composes the {\it covering configuration}
\be
\label{4}
 \xi \equiv \{ \xi_{i\nu} : \; i=1,2,\ldots,N_\nu; \;
\nu = 1,2 \} \; .
\ee

Each thermodynamic state is characterized by the corresponding space of
typical microstates $\mathcal{H}_\nu$ that is a {\it weighted Hilbert
space}, the complete metric space with a weighted norm induced by the
inner product [28]. The latter implies that, if $\{\varphi_n\}$ is
an orthonormal basis in $\mathcal{H}_\nu$ and $f \in \mathcal{H}_\nu$,
so that the inner product $(\varphi,f)$ be given, then the
{\it weighted norm} of $f$ over $\mathcal{H}_\nu$ is
$$
|| f ||_\nu \equiv \sqrt{ \sum_n p_{n\nu} | (\vp_n,f)|^2 } \;   ,
$$
with the set $\{p_{n\nu}\}$ composing a probability measure:
$$
 0 \leq p_{n\nu} \leq 1 \; , \qquad \sum_n p_{n\nu} = 1 \;  .
$$
The statistical average, with a statistical operator $\hat{\rho}_\nu$,
of an operator $\hat{A}$ on $\mathcal{H}_\nu$ is defined as
$$
 \lgl \hat A \rgl_\nu = {\rm Tr}_{\cH_\nu}
\hat\rho_\nu \hat A = \sum_n p_{n\nu} \left (
\vp_n, \hat\rho_\nu \hat A \vp_n \right )  .
$$

The ordered and disordered states are distinguished as follows. There
should exist an {\it order operator} $\hat{\eta}$ such that the averages
\be
\label{5}
\eta_\nu \equiv \lgl \hat\eta \rgl_\nu
\ee
define the order parameters for the corresponding ordered
($\eta_1 \neq 0$) and disordered ($\eta_2 = 0$) states.

For each fixed configuration $\xi$, the Hamiltonian energy operator,
containing the single-atom terms $H_i$ and the binary interaction terms
$H_{ij}$, possesses the representations
\be
\label{6}
\hat H_\nu(\xi) = \sum_i H_i \xi_{i\nu} \; + \;
\sum_{i\neq j} H_{ij} \xi_{i\nu} \xi_{j\nu}
\ee
defined on the related spaces $\mathcal{H}_\nu$. The system as a whole
is described by the Hamiltonian
\be
\label{7}
H(\xi) = \hat H_1(\xi) \bigoplus \hat H_2(\xi)
\ee
defined on the {\it mixture space}
\be
\label{8}
 \cM = {\cal H}_1 \bigotimes {\cal H}_2 \;  .
\ee

The regions of disorder can arise in arbitrary locations and have
different shapes. Therefore, it is necessary to invoke an averaging
procedure over the random configuration set (4). Denoting the
stochastic averaging by the double angle brackets $<< \cdots>>$, one
has the free energy
\be
\label{9}
F = - T\ln \lgl\lgl {\rm Tr} e^{-\bt H(\xi)} \rgl\rgl \; ,
\ee
where the trace is over space (8) and $\beta = 1/T$ is inverse
temperature. This corresponds to annealed disorder, since the disorder
regions are not fixed in space by an external force, but appear in the
system in a self-organized way. The stochastic averaging of an operator
$\hat{A}(\xi)$ over the random configurations $\xi$ can be rigorously
defined [28] as a functional integration over the manifold indicators,
\be
\label{10}
\lgl\lgl \hat A(\xi) \rgl\rgl = \int \hat A(\xi) \;\cD \xi   .
\ee
Physically, the procedure of averaging over the random configurations
is based on the existence of different spatial scales related to atoms
(microscopic scale) and to the regions of disorder (mesoscopic scale).
Different spatial scales also are often connected with different temporal
scales of slow and fast motion [30-33]. Technically, the averaging over
configurations reminds the renormalization procedure [34,35]. All exact
mathematical details of the configuration averaging is given in the review
article [28].

Introducing the {\it effective renormalized Hamiltonian}
\be
\label{11}
\widetilde H \equiv - T \ln \lgl\lgl e^{-\bt H(\xi)} \rgl\rgl
\ee
makes it possible to rewrite the free energy (9) as
\be
\label{12}
 F = - T \ln {\rm Tr} e^{-\bt\widetilde H} \; .
\ee

Thus, the basic point is to find the effective Hamiltonian (11). For the
Hamiltonian representation (6), there has been proved [27,28] the
following statement.

\vskip 2mm

{\it Theorem}. The effective Hamiltonian (11), where $H(\xi)$ is given by
the sum (7) of the Hamiltonian representations (6), takes the form
\be
\label{13}
 \widetilde H = \hat H_1 \bigoplus \hat H_2 \; ,
\ee
in which
\be
\label{14}
 \hat H_\nu = w_\nu \sum_i H_i \; + \; w_\nu^2
\sum_{i\neq j} H_{ij} \; ,
\ee
with the weights $w_\nu$ being the minimizers of the free energy (12),
under the conditions
\be
\label{15}
 w_1 + w_2 = 1 \; , \qquad 0 \leq w_\nu \leq 1 \; .
\ee

\vskip 2mm

The Hamiltonian form (13) is similar to the channel Hamiltonian
representation in scattering theory [36,37].

\section{Double-Well Insulating Lattice}

The detailed derivation of the Hamiltonian for a double-well
insulating lattice has been done in Ref. [15]. It is convenient to
resort to pseudospin representation. To remind the reader the physical
meaning of the pseudospin operators to be used in what follows, let
us recall how they are constructed.

We start with the operators of destruction and creation, $c_{nj}$ and
$c_{nj}^\dagger$, associated with the $n$-th energy level $E_n$ of an
atom in the $j$-th lattice cite. Considering an insulating lattice
implies that atomic transitions between different lattice cites are
suppressed, so that
\be
\label{16}
c_{mi}^\dgr c_{nj} = \dlt_{ij} c_{mi}^\dgr c_{ni} \; .
\ee
A lattice with the unity filling factor is assumed, when the number
of atoms $N$ coincides with the number of lattice cites $N_L$. Then,
limiting the consideration by two lowest energy levels of an atom
in a double-well potential, one has the unipolarity condition
\be
\label{17}
 c_{1j}^\dgr c_{1j} + c_{2j}^\dgr c_{2j} = 1 \; , \qquad
c_{nj} c_{nj} = 0 \; .
\ee

The pseudospin operators are introduced through the transformations
$$
c_{1j}^\dgr c_{1j} = \frac{1}{2} + S_j^x \; , \qquad
c_{2j}^\dgr c_{2j} = \frac{1}{2} - S_j^x \; ,
$$
\be
\label{18}
c_{1j}^\dgr c_{2j} = S_j^z - i S_j^y \; , \qquad
c_{2j}^\dgr c_{1j} = S_j^z + i S_j^y \; .
\ee
The pseudospin operators $S^\alpha_j$ satisfy the standard spin
commutation relations for both the cases, when the operators $c_{nj}$
are either bosonic or fermionic. From the atomic operators, associated
with the energy levels, one can pass to the operators, associated with
the left and right positions of an atom in a double-well potential, by
means of the transformations
\be
\label{19}
c_{1j} = \frac{1}{\sqrt{2}} ( c_{jL} + c_{jR} ) \; , \qquad
c_{2j} = \frac{1}{\sqrt{2}} ( c_{jL} - c_{jR} ) \;  .
\ee
As a result, the pseudospin operators take the form
$$
S_j^x = \frac{1}{2} \left ( c_{jL}^\dgr c_{jR} +
c_{jR}^\dgr c_{jL} \right ) \; , \qquad
S_j^y = -\; \frac{i}{2} \left ( c_{jL}^\dgr c_{jR} -
c_{jR}^\dgr c_{jL} \right ) \; ,
$$
\be
\label{20}
S_j^z = \frac{1}{2} \left ( c_{jL}^\dgr c_{jL} -
c_{jR}^\dgr c_{jR} \right ) \;   .
\ee
The latter representation makes it clear the meaning of the pseudospin
operators, showing that $S^x_j$ describes the atomic tunneling between
the wells of a double-well potential in the $j$-th lattice cite, $S^y_j$
corresponds to the Josephson current between the wells, and the operator
$S^z_j$ defines the atomic imbalance between the left and right wells in
the $j$-the lattice cite.

Employing these pseudospin operators yields [15] the single-atom energy
part
\be
\label{21}
H_i = E_0 - \Om S_i^x \;  ,
\ee
in which $E_0 = (E_1 + E_2)/2$ and $\Omega \approx E_2 - E_1$ is the
interwell tunneling frequency. The binary interactions enter the part
\be
\label{22}
H_{ij} = \frac{1}{2} \; A_{ij} + B_{ij} S_i^x S_j^x -
I_{ij} S_i^z S_j^z \; ,
\ee
with $A_{ij}$ being the direct interaction between atoms in the $i$-th
and $j$-th cites, $B_{ij}$, the interactions of atoms in the process of
their tunneling between the wells, and $I_{ij}$, the interactions of
atoms shifted in one of the wells. All these quantities are explicitly
expressed through the corresponding matrix elements of atomic interaction
potentials over Wannier functions, as is done in Ref. [15].

For creating different states, it is important to have the possibility
of varying atomic interactions in a wide range. In this regard, trapped
atoms in optical lattices provide a unique opportunity. First of all,
nowadays one is able to trap degenerate atomic clouds of a variety of
atomic species possessing different scattering lengths. These are the
whole series of bosonic atoms, $^1$H, $^4$He, $^7$Li, $^{23}$Na, $^{39}$K,
$^{41}$K, $^{52}$Cr, $^{85}$Rb, $^{87}$Rb, $^{133}$Cs, $^{170}$Yb, 
$^{174}$Yb (see the summary in Refs. [1-5,38-44] and references therein) 
and also $^{40}$Ca [45], $^{84}$Sr [46,47], and $^{88}$Sr [48]. There 
are experiments with several degenerate trapped Fermi gases, $^3$He, 
$^6$Li, $^{40}$K, and $^{173}$Yb (see review articles [49-51]). It is 
also possible to create different atomic molecules, made either of bosons, 
$^{23}$Na$_2$, $^{85}$Rb$_2$, $^{87}$Rb$_2$, $^{133}$Cs$_2$, or fermions, 
$^6$Li$_2$, $^{40}$K$_2$, as well as heteronuclear molecules, such as 
Bose-Fermi molecules $^{87}$Rb-$^{40}$K or Bose-Bose molecules 
$^{85}$Rb-$^{87}$Rb and $^{41}$K-$^{87}$Rb [52]. The strength of atomic 
interactions can be varied in a very wide range by means of the Feshbach 
resonance techniques [6,7,38,52,53]. Finally, there are atoms with 
long-range dipolar interactions, such as $^{52}$Cr possessing large 
magnetic moments [54,55], and also Rydberg atoms [56] and polar molecules 
[57]. Moreover, in optical lattices, it is admissible to regulate the shape 
of the double-well potential, varying by this the tunneling frequency [15,58].
Therefore, the system parameters can really be varied in the range of
several orders of magnitude.

Employing the pseudospin representation, discussed above, for the
effective Hamiltonian (14), we obtain
\be
\label{23}
\hat H_\nu = w_\nu N E_0 +
\frac{w_\nu^2}{2} \sum_{i\neq j} A_{ij} \; - \;
w_\nu \Om \sum_j S_j^x \; + \; 
w_\nu^2 \sum_{i\neq j} \left ( B_{ij} S_i^x S_j^x -
I_{ij} S_i^z S_j^z \right ) \;  .
\ee
To some extent, this Hamiltonian is analogous to the effective
Hamiltonian describing solids with pores and cracks [59], in which
the latter are the mesoscopic regions of disorder inside a regular
solid matrix.

For the insulating double-well lattice, the role of the order operator
is played by the pseudospin operator $S^z_j$, whose average gives the
mean atomic imbalance. According to Sec. I, there are two kinds of
averages, one defining the ordered state and another, disordered state.
For an operator $\hat{A}$, these two types of the averages are
\be
\label{24}
\lgl \hat A \rgl_\nu =
\frac{{\rm Tr}_{\cH_\nu}\hat A \exp(-\bt \hat H_\nu)}
{{\rm Tr}_{\cH_\nu}\exp(-\bt\hat H_\nu)} \; ,
\ee
where $\nu = 1,2$. For the order operator, one has
\be
\label{25}
\lgl S_j^z \rgl_1 \neq 0 \; , \qquad
\lgl S_j^z \rgl_2 = 0 \;  .
\ee
This means that the ordered state is characterized by a nonzero order
parameter, while the disordered state, by the zero order parameter.

\section{Mixture Order Parameters}

It is convenient to define the reduced transverse averages
\be
\label{26}
x_\nu \equiv 2 \lgl S_i^x \rgl_\nu \; , \qquad
y_\nu \equiv 2 \lgl S_i^y \rgl_\nu \; ,
\ee
and the normalized order parameter
\be
\label{27}
s_\nu \equiv 2 \lgl S_i^z \rgl_\nu \;  .
\ee

To study the behavior of these averages under varying system
parameters, let us resort to the mean-field approximation. Then, by
using the standard mean-field techniques (see, e.g., [60]), for the
free energy (12), we find
\be
\label{28}
 F = F_1 + F_2 \; , \qquad
F_\nu = \widetilde E_\nu - NT
\ln \left ( 2 \cosh \; \frac{\Lbd_\nu}{2T} \right ) \; ,
\ee
where the first term in $F_\nu$ is
\be
\label{29}
\widetilde E_\nu \equiv w_\nu E_0 N + \frac{w_\nu^2}{2}
\left ( A - B x_\nu^2 + I s^2_\nu \right ) N
\ee
and in the second term, the notations
\be
\label{30}
\Lbd_\nu \equiv w_\nu \sqrt{\Om^2_\nu + w_\nu^2 I^2 s_\nu^2} \; ,
\qquad \Om_\nu \equiv \Om - w_\nu B x_\nu
\ee
are used. The parameters $A,B$, and $I$ are given by the averages
\be
\label{31}
A \equiv \frac{1}{N} \; \sum_{i\neq j} A_{ij} \; , \qquad
B \equiv \frac{1}{N} \; \sum_{i\neq j} B_{ij} \; , \qquad
I \equiv \frac{1}{N} \; \sum_{i\neq j} I_{ij} \;  .
\ee

For the mean tunneling intensity, defined in Eq. (26), we get
\be
\label{32}
 x_\nu = w_\nu \; \frac{\Om_\nu}{\Lbd_\nu} \; \tanh \left (
\frac{\Lbd_\nu}{2T} \right ) \; .
\ee
The average Josephson current is zero,
\be
\label{33}
y_\nu = 0 \;  ,
\ee
as it should be in equilibrium. And for the order parameter (27), we find
\be
\label{34}
 s_\nu = w_\nu^2 s_\nu \; \frac{I}{\Lbd_\nu} \;
\tanh\left ( \frac{\Lbd_\nu}{2T} \right ) \; .
\ee

Let us introduce the dimensionless tunneling frequency and the tunneling
interaction parameter, respectively,
\be
\label{35}
 \om \equiv \frac{\Om}{I+B} \; , \qquad
b \equiv \frac{B}{I+B} \; .
\ee
Also, let us define the dimensionless quantities
\be
\label{36}
\om_\nu \equiv \frac{\Om_\nu}{I+B} = \om - b w_\nu x_\nu \; ,
\qquad
h_\nu \equiv \frac{\Lbd_\nu}{I+B} =
w_\nu \sqrt{\om_\nu^2+(1-b)^2 w_\nu^2 s_\nu^2} \; ,
\ee
corresponding to Eqs. (30). In what follows, let us measure temperature
in units of $I+B$. Then the tunneling intensity (32) reads as
\be
\label{37}
x_\nu = w_\nu \; \frac{\om_\nu}{h_\nu} \; \tanh\left (
\frac{h_\nu}{2T} \right )
\ee
and the order parameter (34) becomes
\be
\label{38}
 s_\nu = w_\nu^2 s_\nu \; \frac{1-b}{h_\nu} \;
\tanh \left ( \frac{h_\nu}{2T} \right ) \;  .
\ee

In view of the definition of the ordered and disordered states,
symbolized in conditions (25), we have
\be
\label{39}
s_1 \neq 0 \; , \qquad s_2 = 0 \;  .
\ee
Therefore for the ordered state, from Eqs. (38) and (39), we get
the tunneling intensity
\be
\label{40}
 x_1 = \frac{\om}{w_1} \;  ,
\ee
while the mean atomic imbalance $s_1$ is defined by the equation
\be
\label{41}
 h_1 = w_1^2 ( 1 - b) \tanh \left ( \frac{h_1}{2T}
\right ) \;  .
\ee

For the disordered state, for which the mean imbalance is zero,
$s_2 = 0$, the tunneling intensity is
\be
\label{42}
x_2 = \tanh \left ( \frac{w_2\om_2}{2T} \right ) \;  .
\ee

In addition to the order parameters $s_1$ and $s_2$, the mixed system
is characterized by the quantities $w_1$ and $w_2$, playing the role of
the geometric weights of the corresponding states. These weights, according
to the theorem of Sec. 2, are defined as the minimizers of the free
energy (12), under the normalization conditions (15). This minimization
yields
\be
\label{43}
w_1 = \frac{2u+\om_1x_1-\om_2x_2}{4u-(1-b)s_1^2} \; , \qquad
w_2 = \frac{2u-\om_1x_1+\om_2x_2-(1-b)s_1^2}{4u-(1-b)s_1^2} \;  ,
\ee
where the notation
\be
\label{44}
u \equiv \frac{A}{I+B}
\ee
is used. The inequality $d^2F/dw_1^2 > 0$ results in the condition
\be
\label{45}
4u + b\left ( x_1^2 + x_2^2 \right ) - (1 - b) s_1^2 > 0 \;  .
\ee
The thermodynamics of the mixed system is described by the set of
Eqs. (39) to (43).

\section{Quantum Phase Transition}

Setting $T=0$ in the above equations gives the tunneling intensities
\be
\label{46}
x_1 = \frac{\om}{w_1} \; , \qquad x_2 = 1
\ee
and the order parameters
\be
\label{47}
 s_1 = \sqrt{ 1 - \; \frac{\om^2}{w_1^2} } \; , \qquad
s_2 = 0 \; .
\ee
The state weights (43) become
\be
\label{48}
w_1 = \frac{2u+b-\om}{4u+2b-1} \; , \qquad
w_2 = \frac{2u+b+\om-1}{4u+2b-1} \;  .
\ee

At zero temperature, the free energy coincides with the internal energy:
\be
\label{49}
 F = \lgl \widetilde H \rgl \qquad ( T = 0 ) \; .
\ee
It is convenient to introduce the dimensionless internal energy
\be
\label{50}
 E(w_1) \equiv
\frac{\lgl\widetilde H \rgl}{N(I+B) } \;  ,
\ee
for which we find
\be
\label{51}
 E(w_1) = e_0 + \frac{1}{4} \left ( 2u + b - 2\om -
\om^2 \right )  - \;\frac{w_1}{2} ( 2u +b - \om) +
\frac{w_1^2}{4} ( 4u + 2b - 1 ) \; ,
\ee
where
\be
\label{52}
e_0 \equiv \frac{E_0}{I+B} \;  .
\ee

The minimization procedure now implies that
\be
\label{53}
 \frac{\prt E(w_1)}{\prt w_1} = 0 \; , \qquad
\frac{\prt^2 E(w_1)}{\prt w_1^2} > 0 \;  .
\ee
The first of these equations leads to Eqs. (48), while the second gives
the inequality
\be
\label{54}
 4u + 2b - 1 > 0 \; .
\ee
By its definition as a matrix element over Wannier functions [15], the
interaction strength $B$ is assumed to be positive and less than $I$.
Then, in view of definition (35), one has
\be
\label{55}
0 < b < \frac{1}{2} \qquad ( 0 < B < I ) \;  .
\ee
Taking into account condition (54) yields
\be
\label{56}
 0 < u_0 < \frac{1}{4} \qquad
\left ( u_0 \equiv \frac{1-2b}{4} \right ) \; .
\ee
In order that $s_1$ in Eq. (47) be real, it is necessary that $\omega$
be less than $w_1$. With $w_1$ from Eq. (48), we have
\be
\label{57}
 \om < \frac{1}{2} \qquad ( s_1 > 0 ) \; .
\ee
Inequalities (54) to (57) specify the region of the system parameters,
when mesoscopic disorder can arise in a double-well optical lattice.

For the double-well lattice with mesoscopic disorder, under the state
weights (48), the internal energy (51) takes the form
\be
\label{58}
 E(w_1) = e_0 + \frac{1}{4} \left ( 2u + b - 2\om - \om^2
\right ) - \; \frac{(2u+b-\om)^2}{4(4u+2b-1)} \; .
\ee
The energy of the mixed state should be compared with that of the
pure states. If the system would be completely ordered, so that
$w_1=1$ while $w_2 = 0$, then the energy would be
\be
\label{59}
 E(1) = e_0 + \frac{1}{4} \left ( 2u + b - \om^2 -1
\right ) \; .
\ee
And, if the system would be absolutely disordered, so that $w_1=0$
while $w_2=1$, then we would have
\be
\label{60}
 E(0) = e_0 + \frac{1}{4} \left ( 2u + b - 2\om - \om^2
\right ) \; .
\ee
From Eqs. (59) and (60), we get
\be
\label{61}
 E(0) - E(1) = \frac{1-2\om}{4} \; ,
\ee
which shows that, under condition (57), the energy of the pure
ordered state is lower than that of the pure disordered state,
that is, the ordered state is preferable, being more stable.

To compare the pure states with the mixed state, we consider the
differences between the related energies,
\be
\label{62}
 E(1) - E(w_1) = (u-u_0)(1-w_1)^2 \; , \qquad
E(0) - E(w_1) = (u-u_0 ) w_1^2 \;  .
\ee
These equations tell us that the quantum phase transition could
occur either at $u =u_0$ or when $w_1=1$. From Eqs. (48) and (56),
we see that the weight $w_1$ can be written as
\be
\label{63}
w_1 = \frac{1}{2} + \frac{1-2\om}{8(u-u_0)} \;  .
\ee
Since, by definition (15), $w_1 \geq 0$ and $w_1 \leq 1$, we find
that
$$
u \geq u_0 \qquad ( w_1 \geq 0) \; ,
$$
\be
\label{64}
 u \geq u_c \qquad ( w_1 \leq 1) \;  ,
\ee
where
\be
\label{65}
 u_c \equiv u_0 + \frac{1-2\om}{4} \; .
\ee
Under condition (57), $u_c > u_0$. Hence, the mixed state exists
only if $u>u_c$.

In this way, the quantum phase transition occurs at $u = u_c$,
given by Eq. (65), when $w_1 = 1$. For $u > u_c$, the mixed state
is more stable, while for $u < u_c$, the pure ordered state becomes
preferable. When $u\ra u_c$, then $w_1\ra 1$. And if $u\ra\infty$,
then $w_1\ra 1/2$.

The parameter $u$, defined in Eq. (44), has the meaning of the
{\it disorder parameter}. For sufficiently large $u > u_c$, the
regions of mesoscopic disorder appear in the system, making this
mixed state preferable as compared to the pure ordered state.

\section{Temperature Phase Transition}

In the pure system without mesoscopic disorder, when $w_1 \equiv 1$,
the order parameter $s_1$, describing the mean imbalance between the
wells of a double-well potential, tends to zero at the temperature
\be
\label{66}
 T_c = \frac{(1-b)\om}{2{\rm arctanh}(\om)} \; .
\ee
But, if the system is mixed, such that $w_1$ is described by Eq. (48),
then the transition temperature, at which $s_1 \ra 0$, is
\be
\label{67}
 T_c^* = \frac{(1-b)\om}{4{\rm arctanh}(2\om)} \; ,
\ee
which is lower than temperature (66). Moreover, the situation for
the mixed lattice is not as simple. The phase transition, at rising
temperature, can happen not as a second-order transition at $T^*_c$
but as a first-order transition at some intermediate temperature
$T_0$ between temperatures (66) and (67), provided that at this
temperature $T_0$ the mixed state becomes less favorable as compared
to the disordered state.

To study the behavior of the mixed state at finite temperatures,
we take into account that the tunneling parameters $\omega$ and $b$
are small, in agreement with inequalities (55) and (57). Hence, for
numerical calculations, these parameters can be neglected. Then Eqs.
(66) and (67) reduce to
\be
\label{68}
T_c = \frac{1}{2} \; , \qquad T_c^* = \frac{1}{8} \;  .
\ee

For convenience, we define the weights as
\be
\label{69}
 w_1 \equiv w \; , \qquad w_2 = 1 - w \; .
\ee
And we solve numerically the equations for the order parameters
\be
\label{70}
 s_1 = \tanh \left ( \frac{w^2 s_1}{2T} \right ) \;, \qquad
w = \frac{2u}{4u-s_1^2} \; .
\ee
These equations, generally, can possess several branches of solutions.
Among them, we have to choose that one corresponding to the smaller
free energy. Also, we need to compare the free energy of the mixed
state with that of the pure system, for which $w\equiv 1$ and the
order parameter is given by the equation
\be
\label{71}
 s_1 = \tanh \left ( \frac{s_1}{2T} \right ) \qquad
( w = 1 ) \; .
\ee
In addition, we have to consider the disordered state, for which
$s_1 = 0$ and $w \ra 1/2$. Comparing the free energies of all
possible solutions, we select that solution whose free energy is
minimal, which corresponds to the most stable system.

In Figs. 3 and 4, we show the behavior of the stable branches
of solutions as functions of temperature for different disorder
parameters $u$. Figure 5 presents the dependence on the disorder
parameter $u$ of the order-disorder transition temperature $T_0$.
The overall picture is as follows.

\vskip 2mm
(i)  Nonpositive disorder parameter:
\be
\label{72}
u \leq 0 \;  .
\ee
Mesoscopic disorder does not appear at all. For temperatures below
the critical $T_c$, the pure ordered state is stable,
\be
\label{73}
  w = 1 \qquad ( 0 \leq T \leq T_c ) \; .
\ee
At $T_c$, the second-order phase transition takes place from the
ordered state, where $s_1 \neq 0$, to the disordered state, where
$s_1 = 0$.

\vskip 2mm
(ii)  In the region of the disorder parameter
\be
\label{74}
 0 < u \leq u_n \;  ,
\ee
where $u_n \approx 0.45$, the pure ordered state remains till the
temperature $T_0$,
\be
\label{75}
w = 1 \qquad ( 0 \leq T \leq T_0 ) \;   ,
\ee
where there happens a first-order phase transition to the disordered
state. The transition temperature is in the interval
\be
\label{76}
 T_c^* < T_0 < T_c \;  ,
\ee
depending on the value of $u$, as is shown in Fig. 5.

\vskip 2mm
(iii)  In the region of the parameter
\be
\label{77}
 u_n < u < \frac{1}{2} \; ,
\ee
there exists the {\it nucleation temperature}
\be
\label{78}
 T_n = \frac{\sqrt{2u}}{2{\rm arctanh}\sqrt{2u}} \; ,
\ee
below which the pure ordered state is stable,
\be
\label{79}
w = 1 \qquad ( 0 \leq T \leq T_n ) \;  .
\ee
But, at $T_n$, there appears mesoscopic disorder, and the system
becomes mixed,
\be
\label{80}
 w < 1 \qquad ( T_n < T \leq T_0 ) \; .
\ee
At $T_0$, such that
\be
\label{81}
 T_n < T_0 < T_c \; ,
\ee
there occurs a first-order phase transition from the mixed state to
the disordered state. When $u \ra u_n$, then $T_n \ra T_0$.

\vskip 2mm
(iv)  For the parameter
\be
\label{82}
 u = \frac{1}{2} \;  ,
\ee
the nucleation temperature $T_n = 0$. At zero temperature, the system
is purely ordered,
\be
\label{83}
 w = 1 \qquad ( T = 0 ) \; .
\ee
But it is mixed for all temperatures above zero up to $T_0$,
\be
\label{84}
 w < 1 \qquad ( 0 < T \leq T_0 ) \;  .
\ee
At $T_0$, there is a first-order phase transition from the mixed state
to the disordered state.

\vskip 2mm
(v)  In the region
\be
\label{85}
 \frac{1}{2} < u < \frac{3}{2} \;  ,
\ee
the system is always mixed,
\be
\label{86}
  w < 1 \qquad ( 0 \leq T \leq T_0 ) \; ,
\ee
up to the temperature $T_0$, where a first-order phase transition
happens.

\vskip 2mm
(vi)  For the value
\be
\label{87}
 u = \frac{3}{2} \;  ,
\ee
the system is mixed up to the temperature $T_c^*$,
\be
\label{88}
 w < 1 \qquad ( 0 \leq T \leq T_c^* ) \;  .
\ee
The temperature $T_c^*$ corresponds to a tricritical point that is
the end-point on the line of first-order phase transitions, separating
these from the line of second-order transitions.

\vskip 2mm
(vii)  At large values of the disorder parameter, such that
\be
\label{89}
  u > \frac{3}{2} \; ,
\ee
the system is always mixed,
\be
\label{90}
 w < 1 \qquad ( 0 \leq T \leq T_c^* ) \;  .
\ee
At temperature $T_c^*$, there occurs a second-order phase transition
to the disordered state, where $s_1 = 0$.

This analysis shows that the regions of mesoscopic disorder arise
in a double-well lattice when the disorder parameter $u$ is
sufficiently large. By its definition, the parameter $u$ is the
ratio of the disordering interactions to the effectively ordering
ones. The appearance of the mesoscopic disorder is due to the
competition between the ordering and disordering interactions.

\section{Conclusion}

The double-well optical lattice in the insulating state has been
considered. In such a lattice, depending on the system parameters,
there can exist the ordered state, when the mean atomic imbalance
is nonzero and the disordered state, with the zero atomic imbalance.

We have shown, that, in addition to these pure states, there can
arise an intermediate, mixed, state, when, inside a generally
ordered lattice, there develop the regions of mesoscopic disorder.
These regions are mesoscopic in the sense of their typical sizes
being intermediate between the lattice spacing and the total system
size. The shapes and locations of the mesoscopic regions of disorder
are random, which requires to invoke the averaging over all their
random spatial configurations. Accomplishing this averaging allows
one to define an effective renormalized Hamiltonian and to calculate
all thermodynamic characteristics. All mathematical details, 
describing the configuration averaging have been given in the earlier 
papers [21-28,61,62].
 
The considered mesoscopic disorder appears in a self-organized
way and may happen without any external random forces. The stability
of this kind of disorder is due to the existence in the system of
different types of pair interactions, among which there should exist
the interactions that do not favor the ordering. The ratio of the
competing disordering to ordering interactions plays the role of a
disorder parameter. The latter has to be sufficiently large in order
to stabilize the mixed state with the mesoscopic disorder.

As is discussed in Sec. 3, the optical lattices of trapped atoms
provide a unique opportunity of regulating practically all system
parameters. This can be done by varying the types of atoms, by
creating atomic molecules, and by changing the lattice parameters.
It is also possible to regulate the shape of the double-well potential,
varying by this the tunneling frequency [15,58]. The strength of
atomic interactions can be varied in a very wide range by means of
the Feshbach resonance techniques [6,7,38,52,53]. Some atoms possess
interactions with long-range dipolar forces, such as $^{52}$Cr [54,55]
and Rydberg atoms [56]. There are also polar molecules with long-range
forces [57]. It is admissible to create molecules of large size, whose
interactions also have long-range nature, such as the interactions in
liquids and plasmas [63], having, for instance, the Lennard-Jones
form [64].

One more possibility of influencing the system states is by invoking
external fields. Pumping energy into the system can produce effects
that are similar to increasing the system temperature. It is known
[65,66] that sometimes even weak external temporal perturbations
essentially change the system behavior. The stationary action of
alternating external fields can be described as the effective rise
of temperature, which can lead to the appearance of mixed states,
as in the statistical approach to quantum turbulence [67]. It may 
also be possible to excite a part of atoms out of the bound states
and to create different asymmetric localized states above the 
double-well lattice, such as exist in bichromatice lattices [68].
The localization should occur in the disordered spacial regions.

In order to give an understanding of how the effective temperature
can be estimated, in the case of a stationary action of an external
alternating potential, let us give an example. Suppose the system
is subject to the action of an external modulating potential $V(t)$,
with amplitude $V_{mod}$, alternating with a period $t_{mod}$ during 
the time $\Delta t$. If the modulation period is much longer than the 
local equilibration time, then the system is in quasi-equilibrium at 
each instant of time. If there exists a thermostate with temperature $T$
and, in addition, the modulating potential pumps energy into the
system, then the effective temperature can be estimated as
$$
  T_{eff} = T + T_{mod} \; ,
$$
where the modulation temperature is
$$
  T_{mod} = \frac{\Delta t}{t_{mod}} V_{mod} = 
\frac{V_{mod}}{2 \pi} \omega_{mod} \Delta t \; .
$$

The observation of the arising mesoscopic disorder and the level of 
disordering can be realized by optical measurements [69] or scattering 
experiments [70]. 

In this way, there are many possibilities to vary the system
parameters. Therefore, it seems feasible to find the conditions
favoring the appearance of the mesoscopic disorder. The level of
disorder can also be regulated, which can be used in applications.

\newpage

\begin{center}
{\Large {\bf Figure Captions}}
\end{center}

\vskip 3cm

{\bf Fig. 1}  Scheme of the ordered state in an insulating double-well
optical lattice.

\vskip 1cm
{\bf Fig. 2}  Scheme of the intermediate mixed state, with mesoscopic
disordered regions inside the ordered matrix denoted by arrows.

\vskip 1cm
{\bf Fig. 3}  Imbalance order parameter as a function of temperature (in
dimensionless units), with the marked points of the order-disorder phase
transitions, for different disorder parameters: (1) $u = 0$; (2) $u = 0.1$;
(3) $u = 0.3$; (4) $u = 0.4$; (5) $u = 0.51$; (6) $u = 0.75$; (7) $u = 1$;
(8) $u = 1.5$.

\vskip 1cm
{\bf Fig. 4}  Geometric weight of the ordered phase in the mixed system
as a function of temperature (in dimensionless units), with the marked
points of the order-disorder phase transitions, for the same disorder
parameters as in Fig. 3.

\vskip 1cm
{\bf Fig. 5}  Order-disorder transition temperature (in dimensionless
units) as a function of the disorder parameter $u$. The phase transition
is of second order for $u = 0$ and $u > 3/2$; it is of first order for
$0 < u < 3/2$; and $u =3/2$ corresponds to a tricritical point.

\newpage

\begin{figure}[h]
\centerline{\includegraphics[width=14cm]{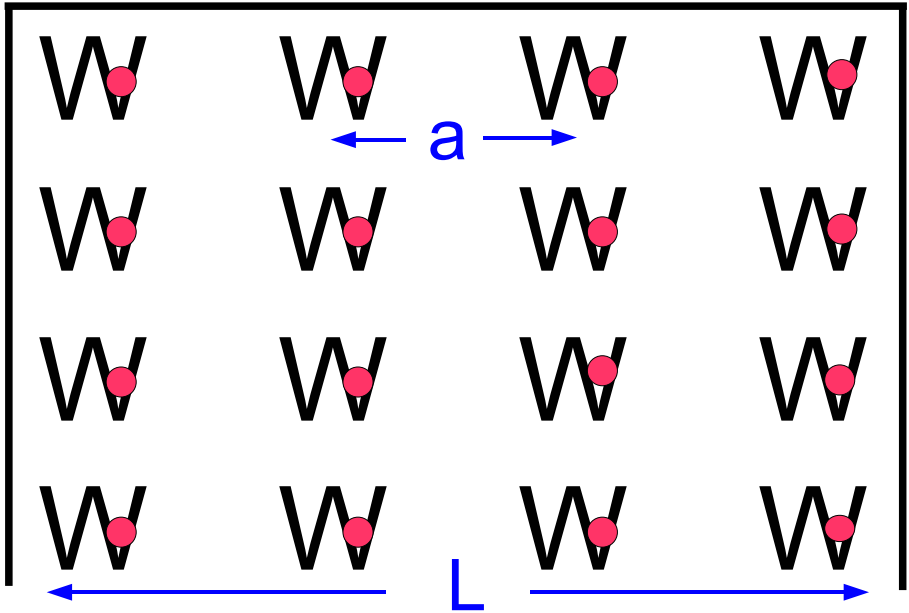}}
\caption{Scheme of the ordered state in an insulating double-well
optical lattice.}
\label{fig:Fig.1}
\end{figure}

\newpage
\begin{figure}[h]
\centerline{\includegraphics[width=14cm]{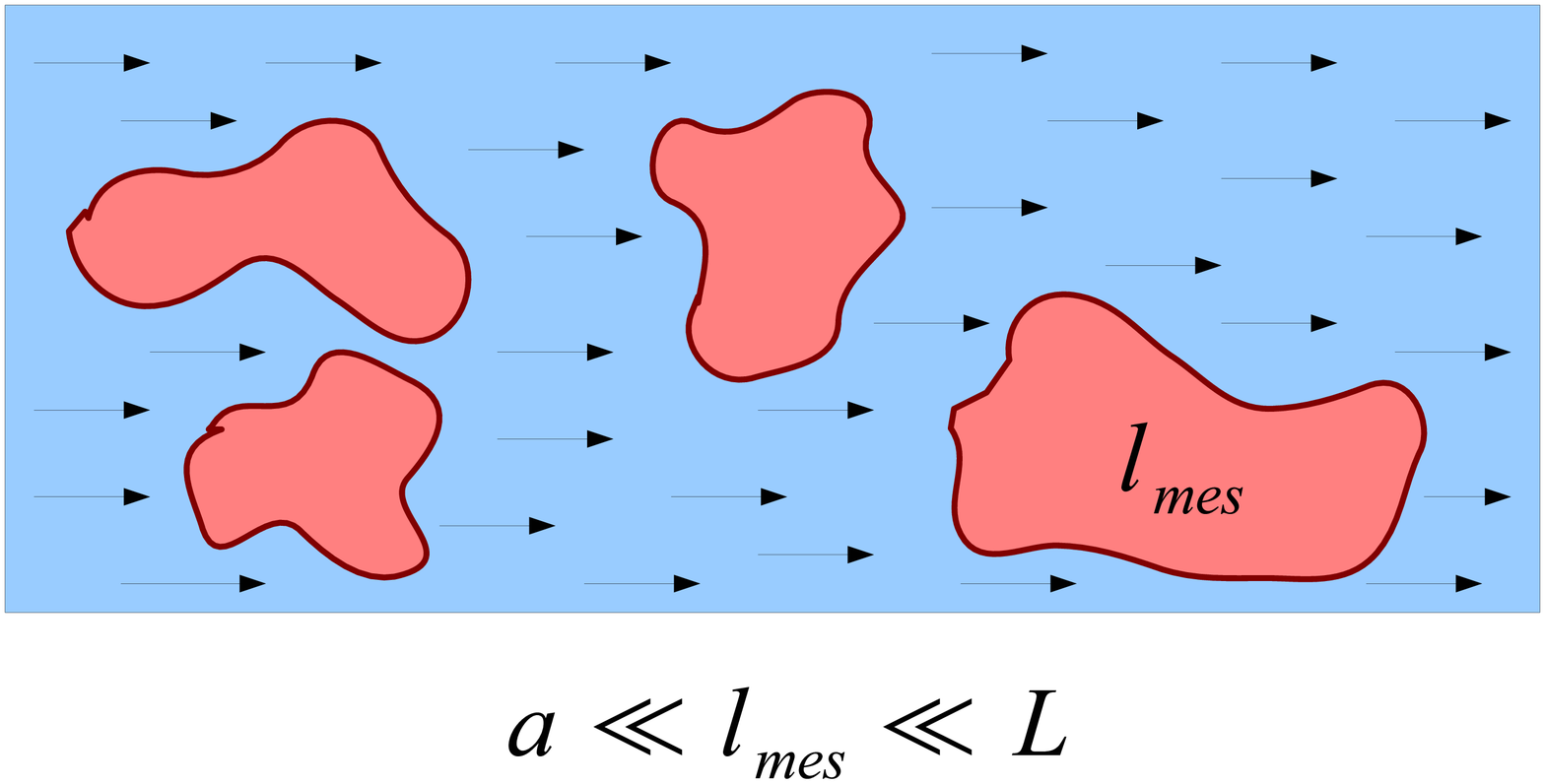}}
\caption{Scheme of the intermediate mixed state, with mesoscopic
disordered regions inside the ordered matrix denoted by arrows.}
\label{fig:Fig.2}
\end{figure}

\newpage
\begin{figure}[h]
\centerline{\includegraphics[width=14cm]{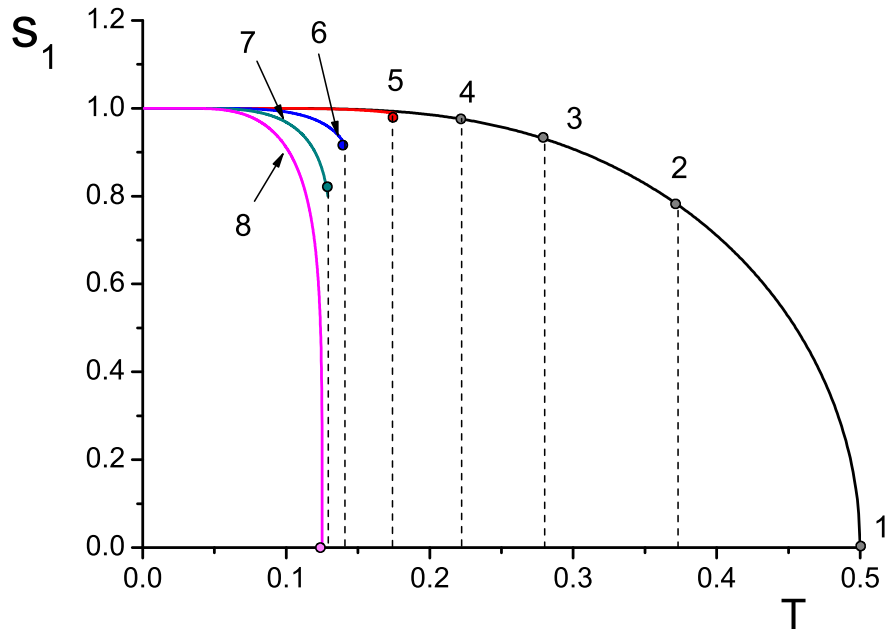}}
\caption{Imbalance order parameter as a function of temperature (in
dimensionless units), with the marked points of the order-disorder phase
transitions, for different disorder parameters: (1) $u = 0$; (2) $u = 0.1$;
(3) $u = 0.3$; (4) $u = 0.4$; (5) $u = 0.51$; (6) $u = 0.75$; (7) $u = 1$;
(8) $u = 1.5$.}
\label{fig:Fig.3}
\end{figure}

\newpage
\begin{figure}[h]
\centerline{\includegraphics[width=14cm]{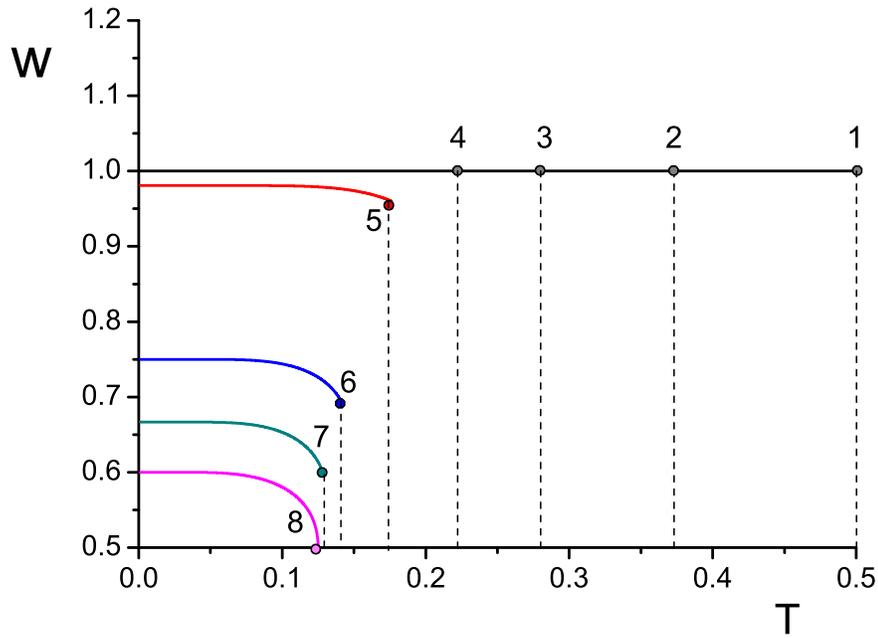}}
\caption{Geometric weight of the ordered phase in the mixed system
as a function of temperature (in dimensionless units), with the marked
points of the order-disorder phase transitions, for the same disorder
parameters as in Fig. 3.}
\label{fig:Fig.4}
\end{figure}

\newpage
\begin{figure}[h]
\centerline{\includegraphics[width=14cm]{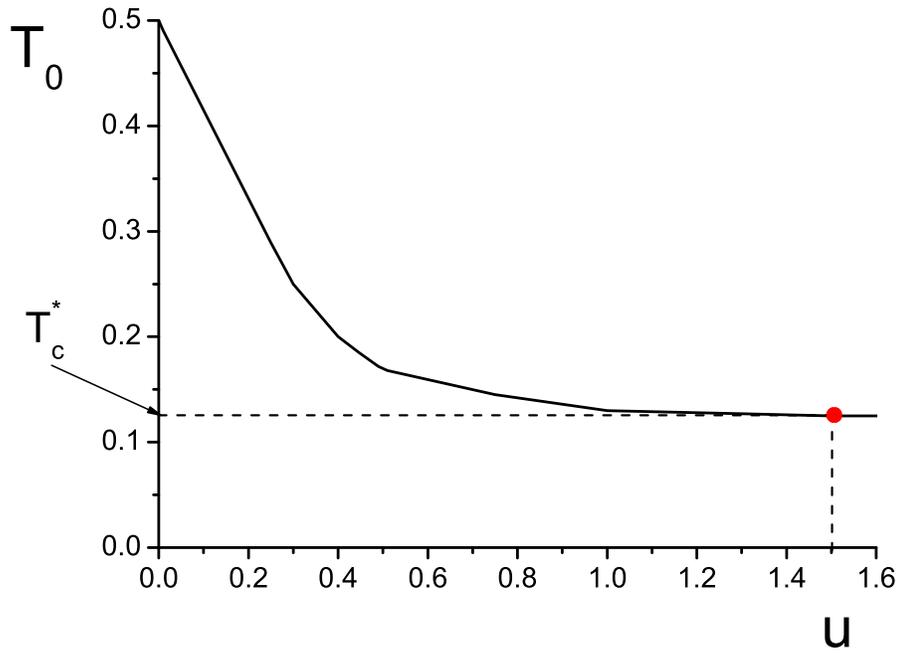}}
\caption{Order-disorder transition temperature (in dimensionless
units) as a function of the disorder parameter $u$. The phase transition
is of second order for $u = 0$ and $u > 3/2$; it is of first order for
$0 < u < 3/2$; and $u =3/2$ corresponds to a tricritical point.}
\label{fig:Fig.5}
\end{figure}

\end{document}